# Ultrafast laser writing quill effect in low loss waveguide fabrication regime

JUN GUAN,[1,3] XIANG LIU,[1] MARTIN J. BOOTH [1,2,*]

[1] Department of Engineering Science, University of Oxford, Parks Road, Oxford, OX1 3PJ, UK
[2] Centre for Neural Circuits and Behaviour, University of Oxford, Mansfield Road, Oxford, OX1 3SR, UK
[3] jun.guan@eng.ox.ac.uk
*Corresponding author: martin.booth@eng.ox.ac.uk

**Quill effect, one of the intriguing phenomena in ultrafast laser writing, to our knowledge, has not been studied in low-loss-waveguide (LLW) writing regime yet, probably due to its 'invisibility' under conventional white-light microscope in that regime. In this report, with help of adaptive third harmonic generation microscopy we reveal the quill effect in LLW writing regime and study its influences on the properties of written photonic integrated components in term of polarization-related properties in fused silica and beam-splitting ratio of three-waveguide-coupler in borosilicate glass.**

Quill effect [1], which manifests as a change in material modification by reversing the writing direction of ultrafast laser direct writing, has been actively studied over the last decade since it was reported by Poumellec et al.[2]. Kazansky et al. suggested quill effect arises from pulse front tilt (PFT) of the writing beam [1], which is supported by following works [3-5] and is improved by Salter et al. by proving that both PFT and time-invariant focal asymmetry may give rise to quill effect [6]. So far all prior works predominately focused on quill effect arising in laser irradiated regions in fused silica that exhibited damage-like textures and were written well above low-loss-waveguide (LLW) writing regime [1-9]. In that damage-resulting writing regime and material, quill effect was clearly visible under a transmission mode microscope but this fabrication regime is not particularly useful for function photonic components. From the prospective of practical application, understanding quill effect in LLW writing regime and in both fused silica and another commonly used glass-borosilicate glass [10-14]- is very important for the design and fabrication of femtosecond-laser-written photonic integrated circuit (FLW-PIC), since some high-demanding applications like polarization-sensitive quantum information processing based on photon [11-15] impose strict requirements on the polarization properties of the fabricated FLW-PIC components, which quill effect may have influence on.

In this letter, by means of adaptive third harmonic generation (THG) microscopy [16,17], we 'visualize' quill effect in LLW writing regime and study the influences of quill effect on the polarization-related properties of low loss waveguide in fused silica as well as on the beam-splitter-ratio of three-waveguide-coupler (tritter) in borosilicate glass.

The laser, which was employed to write FLW-PIC components inside glass, was the second harmonic of a regenerative amplified Yb:KGW laser (Light Conversion Pharos SP-06-1000-pp) with 1 MHz repetition rate, 514 nm wavelength, 170 fs pulse duration laser. The power of the laser beam was regulated through combination of a motorized rotating half waveplate and a polarization beam splitter before being circularly-polarized and focused 120 μm below the top surface of the glass chip with a 0.5 NA objective lens. The glass chip, which was fixed on a three-axis air bearing stage (AerotechABL10100L/ABL10100L/ANT95-3-V), was transversely scanned relative to the focus to inscribe FLW-PIC components. The custom-built adaptive THG microscope was as described in [16]. Aberration was corrected through sequentially adjusting the amplitudes of the Zernike polynomial modes added to the deformable mirror, in order to maximize an image quality metric, which was defined as the total image intensity. Both system aberrations resulted from the microscope light path and the sample-introduced aberrations were corrected. We note here that while the THG depends upon the third-order non-linear optical properties of the material, in practice the measured THG signal correlates spatially to changes in refractive-index profiles[18, 17], such that a rapid spatial transition between different refractive indices typically correlates with a high signal. During light-coupling test, 780 nm-wavelength laser beam from a polarization-maintaining fiber-coupled laser source (Thorlabs S1FC780PM) was butt-coupled into waveguide through a polarization-maintaining fiber (Thorlabs P1-630PM-FC-5); a photodiode power sensor was used to measure the power. The polarization states of the input and output beams into and from a waveguide were measured with a polarimeter (Polarization Analyzer SK010PA-VIS,

Schäfter + Kirchhoff GmbH). Propagation losses were measured through the cutback approach.

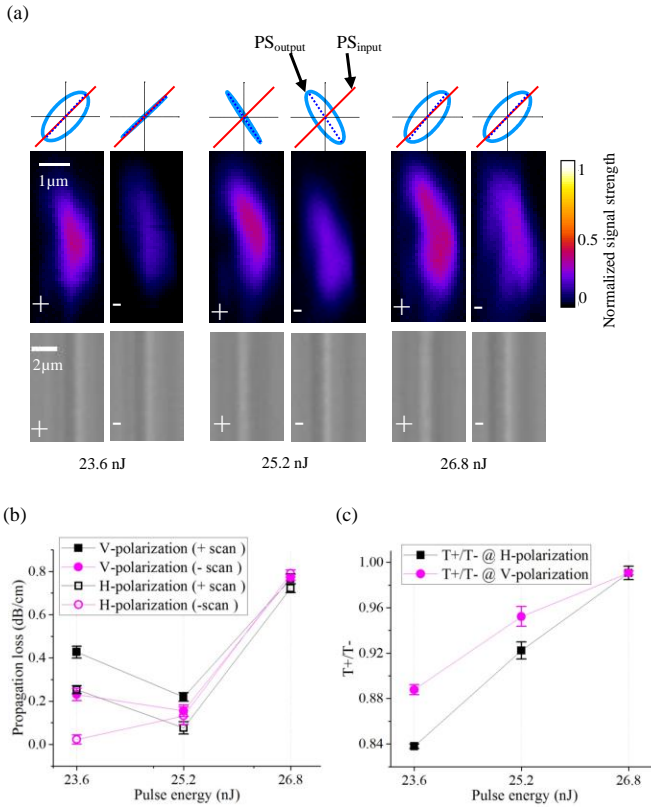

**Fig. 1** Quill effect in LLW writing regime in fused silica. (a) Cross-sectional THG images of the three pairs of waveguide are shown in the middle row (in false color), which were taken with the same THG microscope settings; '+' and '-' denote opposite scan directions; measured output polarization state (PS$_{output}$, represented by polarization state ellipse) with respect to that of input (PS$_{input}$) is on top of each THG image; below each THG image is the corresponding longitudinal white light microscopy image. (b) Measured propagation loss at horizontal (H-polarization) and vertical (V-polarization) polarization; the error bars represent peak-to-valley variation of multiple tests. (c) Ratio between measured throughputs of a pair of waveguides at horizontal and vertical polarization; T+ and T- represent throughputs of waveguide written in opposite scan directions.

In fused silica (LEONI SQ0), three pairs of waveguides were written with the same scan speed of 10 μm/s and sample-surface-level pulse energy of 23.6, 25.2 and 26.8 nJ respectively; within each pair, two waveguides were written the same writing parameters except opposite scan directions. All the waveguides were written in the same chip to reduce the influences of writing system drift and material non-uniformity, as well as to guarantee the same waveguide length for all the waveguide. Their cross-sectional THG images are shown in Fig. 1(a), where we can see that for the pair written with pulse energy of 23.6 nJ, both the profiles and signal strengths of their THG images were clearly different, which contrasted with the similarity of their transmission microscopy images in Fig. 1(a); at pulse energy of 25.2 nJ, their signal strengths were still different but their profiles got similar; as the pulse energy was increased to 26.8 nJ, both their signal strengths and profiles became more similar. This qualitative THG trend correlated with following four measured quantitative trends: I:, polarization modulation, which is represented by the measured polarization-state ellipse of the output light (PS$_{output}$) with respect to that of the input light (PS$_{input}$). as shown in the top row figures of Fig. 1(a); PS$_{input}$ was kept unchanged during measurements of all waveguides; II:, propagation losses in horizontal and vertical polarization directions as shown in Fig. 1(b); III:, the throughput ratio T+/T- at horizontal and vertical polarization directions in Fig. 1(c) T+ and T- denote throughputs of waveguides written in opposite directions within a pair; IV:, the near field mode profiles waveguide at horizontal and vertical polarization of input light as shown in Fig. 2; in Fig. 2 (b), Dx/Dy is the ratio between the diameters of a mode profile along x-axis (Dx) and y-axis (Dy).

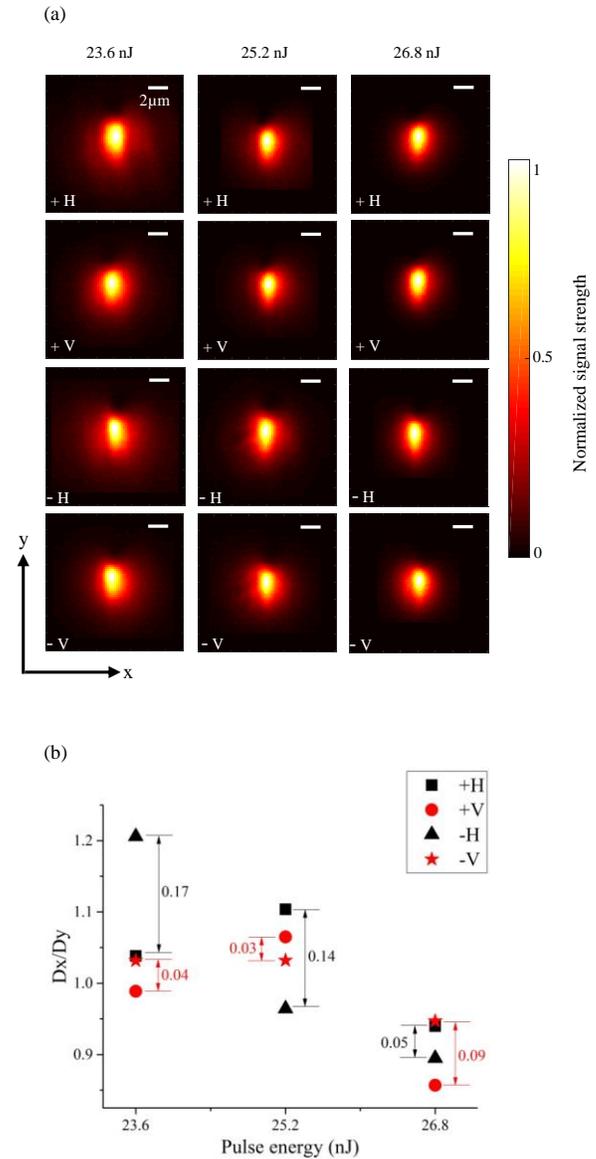

**Fig. 2** Near field mode profiles of waveguides at horizontal and vertical polarized input light. (a) Recorded near field mode

profile images; images were intensity-normalized and in false color; all scale-bars represent 2 μm. (b) Dx/Dy of waveguides; Dx and Dy are diameters of near field of mode profiles in x- and y- axis, under $1/e^2$ intensity definition; marked numbers are the Dx/Dy difference between waveguide written in opposite directions and tested in orthogonal polarizations.

As one can see from Fig. 1 and Fig. 2, the optical properties of waveguides, whose variations correlate to that of THG images under quill effect, are polarization related. To understand these correlations and quill effect in the LLW regime, we start from the origins of polarization-dependency of those waveguides. There are three known types of birefringence in femtosecond laser written waveguides in fused silica [19]: shape birefringence [20], stress-induced birefringence [21] and birefringence induced by nanograting structures [22]. It has been proven that circular polarization [23] of writing beam or repetition rates over 500 kHz [24], which both are the case in this report, can be used to avoid the generation of nanograting. Therefore, it is reasonable that we only consider shape and stress-induced birefringence in the waveguides shown in Fig. 1. Obviously shape birefringence results from the noncircular cross-sectional shape of waveguide; but stress-introduced birefringence also originates from the noncircular or non-uniform cross-sectional profile of waveguide considering the isotropic properties of glass before being irradiated with femtosecond laser. Both shape and stress-introduced birefringence in laser writing single mode waveguide are similar to that of the single mode fiber, in term of birefringence formation mechanisms. Therefore, based on the study on birefringence of single mode fiber [25, 26], both shape birefringence and stress-introduced birefringence of waveguides depend on their profiles that are correlated to their THG image [17]. Based on the relationship between propagation loss, coupling loss and throughput at orthogonal polarizations of input light, as well the relationship between refractive index change and near filed mode diameters at orthogonal polarizations [19], the correlations between THG images shown in Fig. 1 and the optical properties shown in Fig. 1 and Fig. 2 can be understood.

In borosilicate glass, another broadly used FLW-PIC substrate especially for applications like quantum information processing [11-14, 27-29], to our knowledge, there is no quill effect reported in it yet, in particular at MHz repetition rate of write laser. This is due to that its softening temperature is lower than that of fused silica and consequently thermal effect [30] can easily make the quill effect 'invisible' under a conventional transmission microscope. However, we have observed that a difference is visible under the adaptive THG microscope. To clearly reveal the quill effect in borosilicate glass, writing parameters of 86 ~ 113 nJ pulse energy and 0.3 mm/s scan speed were selected to fabricate waveguide in Corning EAGLE 2000. The cross-sectional profile of a waveguide is shown by its THG images in Fig. 3(a), in which the THG signal in the region at the bottom is much stronger than that in the upper parts of the waveguide. The region with strong THG signal is believed due to micro-explosion [28]. As indicated in Fig. 3(b), six waveguides were written with alternate scan directions, pulse energy of 113 nJ, scan speed of 0.3 mm/s and separation of 4 μm in x-axis; the THG image of their bottom region is shown in Fig. 3(c), from which one can see the clear quill effect manifested through THG signal strength contrast of waveguides written in opposite directions.

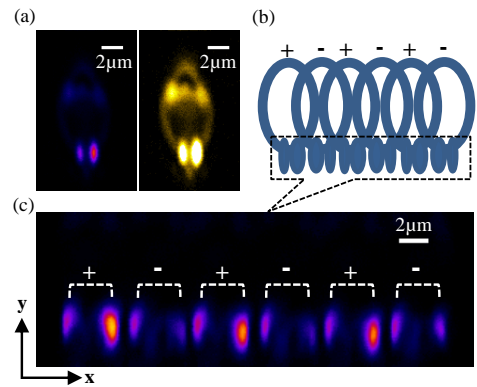

**Fig. 3** Quill effect in Corning EAGLE 2000 glass. (a) Cross-sectional THG images (in false colors) of waveguide written with pulse energy of 86 nJ, scan speed of 0.3 mm/s; the two THG images are of the same waveguide but in different colormaps to emphasis the bottom region (left) and reveal the full profile (right) respectively. (b) Schematic of waveguide array written with alternate scan directions. (c) THG image of the bottom region of the waveguide array written with pulse energy of 113 nJ, scan speed of 0.3 mm/s.

The quill effect has also been observed, to a lesser extent, in FLW-PIC components in borosilicate glass. The effect normally is not as striking as that shown in Fig. 3, but it is still visible under the THG microscope and more importantly it has significant impact on the property of some FLW-PIC component. As shown Fig. 4, from a three-dimensional tritter, which was written in Corning EAGLE 2000 glass with our optimized writing parameter for this kind of component: 85 nJ pulse energy and 2 mm/s scan speed, one can see the quill effect in particular at the lower regions (highlighted by dot-dashed rectangles) of the three coupling waveguides W1, 2 and 3. More importantly this quill effect will affect the beam-splitting ratio of the tritter due to the fact that quill effect will effectively result in a shift of the as-built relative positions between the three coupling waveguides compared with the design values, as we demonstrated in reference [31]. Since the output beam-splitting ratio of a tritter is very sensitive to the relative positions between three coupling waveguides at its coupling region.

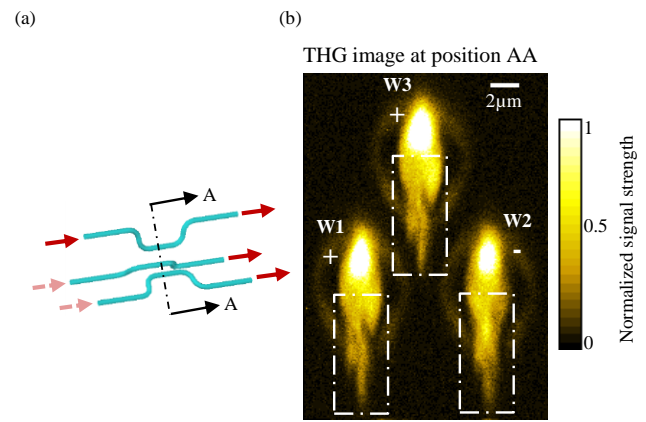

**Fig. 4** A tritter in EAGLE 2000 glass. (a) Schematic of tritter. (b) Cross-sectional THG image of a tritter (in false color) written with pulse energy of 85 nJ and scan speed of 2 mm/s; W1, 2, 3

denote waveguide 1, 2, 3, which were written with alternate scan directions represented by '+' and '-'

In conclusion, with the help of adaptive THG microscopy, we revealed the quill effect and studied its influence on the performances of the fabricated FLW-PIC components in low-loss-waveguide writing regime, both in fused silica and borosilicate glass. This study not only gives us new phenomenological insight into quill effect in previously un-explored regime, but more importantly the information provided is critical for fabricating FLW-PIC components for some high-demanding applications like polarization-related quantum information processing. Quill effect can be reduced through control the spatial intensity distribution of writing beam demonstrated by Salter et al [6], or simply be avoided by using the same scan direction.

**Funding.** UK Engineering and Physical Sciences Research Council (EPSRC) EP/M013243/1 and EP/K034480/.

## References

1. P. Kazansky, W. Yang, E. Bricchi, J. Bovatsek, A. Arai, Y. Shimotsuma, K. Miura, and K. Hirao, Appl. Phys. Lett. **90**(15), 151120 (2007).
2. B. Poumellec, L. Sudrie, M. Franco, B. Prade, A. Mysyrowicz, Opt. Express **11** (9), 1070 (2003).
3. W. Yang, P. G. Kazansky, Y. Shimotsuma, M. Sakakura, K. Miura, and K. Hirao, Appl. Phys. Lett. **93**(17), 171109 (2008).
4. D. N. Vitek, E. Block, Y. Bellouard, D. E. Adams, S. Backus, D. Kleinfeld, C. G. Durfee, and J. A. Squier, Opt. Express **18**(24), 24673–24678 (2010).
5. A. Patel, Y. Svirko, C. Durfee, and P. G. Kazansky, Sci. Rep **7**, 12928 (2017).
6. P. S. Salter and M. J. Booth, Appl. Phys. Lett. **101**, 141109 (2012).
7. Y. Bellouard and M.-O. Hongler, Opt. Express **19**(7), 6807 (2011)
8. C. Corbari, A. Champion, M. Gecevičius, M. Beresna, Y. Bellouard, and P. G. Kazansky, Opt. Express **21**(4), 3946 (2013).
9. H. Song, Y. Dai, J. Song, H. Ma, X. Yan, and G. Ma, Appl. Phys. A **123**, 255 (2017).
10. D. Tan, K. N. Sharafudeen, Y. Yue, and J. Qiu, Prog. Mater. Sci **76**, 154 (2016).
11. A. Crespi, R. Ramponi, R. Osellame, L. Sansoni, I. Bongioanni, F. Sciarrino, G. Vallone, and P. Mataloni, Nat. Commun. **2**, 566 (2011).
12. L. Sansoni, F. Sciarrino, G. Vallone, P. Mataloni, A. Crespi, R. Ramponi, and R. Osellame, Phys. Rev. Lett. **108**, 010502 (2012).
13. G. Corrielli, A. Crespi, R. Geremia, R. Ramponi, L. Sansoni, A. Santinelli, P. Mataloni, F. Sciarrino, and R. Osellame, Nat. Commun. **5**, 4249 (2014).
14. I. Pitsios, L. Banchi, A. S. Rab, M. Bentivegna, D. Caprara, A. Crespi, N. Spagnolo, S. Bose, P. Mataloni, R. Osellame, and F. Sciarrino, Nat. Commun. **8**, 1569 (2017).
15. J. Zeuner, A. N. Sharma, M. Tillmann, R. Heilmann, M. Gräfe, A. Moqanaki, A. Szameit, and P. Walther, npj Quantum Inf. **4**, 13 (2018).
16. A. Jesacher, A. Thayil, K. Grieve, D. Débarre, T. Watanabe, T. Wilson, S. Srinivas, and M. Booth, Opt. Lett. **34**, 3154 (2009).
17. G. D. Marshall, A. Jesacher, A. Thayil, M. J. Withford, and M. Booth, Opt. Lett. **36**, 695 (2011).
18. Müller, Squier, Wilson, and Brakenhoff, J. Microsc **191**, 266 (1998).
19. J. Guan, X. Liu, P. S. Salter, and M. J. Booth, Opt. Express **25**, 4845 (2017).
20. L. B. Jeunhomme, *Single-Mode Fiber Optics Principles and Applications* (Marcel Dekker, Inc 1990).
21. V. R. Bhardwaj, P. B. Corkum, D. M. Rayner, C. Hnatovsky, E. Simova, and R. R. Taylor, Opt. Lett. **29**(12), 1312 (2004).
22. Y. Shimotsuma, P. G. Kazansky, J. Qiu, K. Hirao, Phys. Rev. Lett. **91**(24), 247405 (2003).
23. R. S. Taylor, E. Simova, and C. Hnatovsky, Opt. Lett.s **33**(12), 1312 (2008).
24. W. Yang, E. Bricchi, P. G. Kazansky, J. Bovatsek, A. Y. Arai, Opt. Express **14**(21), 10117 (2006).
25. J. Sakai and T. Kimura, IEEE J. Quant. Electon. **QE-17**, 1041 (1981).
26. J. Sakai and T. Kimura, IEEE J. Quant. Electon. **QE-18**, 1899 (1982).
27. A. Crespi, R. Osellame, R. Ramponi, V. Giovannetti, R. Fazio, L. Sansoni, F. De Nicola, F. Sciarrino, and P. Mataloni, Nat. Photonics **7**, 322 (2013).
28. A. Crespi, R. Osellame, R. Ramponi, D. J. Brod, E. F. Galvão, N. Spagnolo, C. Vitelli, E. Maiorino, P. Mataloni, and F. Sciarrino, Nat. Photonics **7**, 545 (2013).
29. T. Giordani, F. Flamini, M. Pompili, N. Viggianiello, N. Spagnolo, A. Crespi, R. Osellame, N. Wiebe, M. Walschaers, A. Buchleitner, and F. Sciarrino, Nat. Photonics **12**, 173 (2018).
30. S. M. Eaton, H. Zhang, M. L. Ng, J. Li, W.-J. Chen, S. Ho, and P. R. Herman, Opt. Express **16**, 9443 (2008).
31. J. Guan, X. Liu, A. J. Menssen, M. J. Booth, https://arxiv.org/abs/1802.08016.